\documentclass[journal]{IEEEtran}
\usepackage{amsthm}
\usepackage{amsmath}
\usepackage{amssymb}
\usepackage{graphicx}
\usepackage{url}
\usepackage{dsfont}
\usepackage{float}
\usepackage{bm}
\usepackage{ifthen}
\usepackage[usenames,dvipsnames]{color}
\usepackage{mathrsfs}
\usepackage[colorlinks=true,citecolor=blue,urlcolor=black]{hyperref}
\usepackage{float}

\newcommand{\tn}[1]{\textnormal{#1}}
\newcommand{\be}{\begin{equation}}
\newcommand{\ee}{\end{equation}}

\newcommand{\sket}[1]{{\ensuremath{\lvert#1\rangle}}}
\newcommand{\lket}[1]{{\ensuremath{\left\lvert#1\right\rangle}}}
\newcommand{\ket}[1]{\if@display\lket{#1}\else\sket{#1}\fi}

\newcommand{\sbra}[1]{{\ensuremath{\langle#1\rvert}}}
\newcommand{\lbra}[1]{{\ensuremath{\left\langle#1\right\rvert}}}
\newcommand{\bra}[1]{\if@display\lbra{#1}\else\sbra{#1}\fi}

\newcommand{\sbraket}[2]{{\ensuremath{\langle#1\rvert#2\rangle}}}
\newcommand{\lbraket}[2]{{\ensuremath{\left\langle#1\!\left\rvert\vphantom{#1}#2\right.\!\right\rangle}}}
\newcommand{\braket}[2]{\if@display\lbraket{#1}{#2}\else\sbraket{#1}{#2}\fi}

\newcommand{\sketbra}[2]{{\ensuremath{\lvert #1\rangle\!\langle #2\rvert}}}
\newcommand{\lketbra}[2]{{\ensuremath{\left\lvert #1\right\rangle\!\!\left\langle #2\right\rvert}}}
\newcommand{\ketbra}[2]{\if@display\lketbra{#1}{#2}\else\sketbra{#1}{#2}\fi}

\ifCLASSINFOpdf
\else
\fi

\begin{document}
\title{Random Variation of Detector Efficiency: A Countermeasure against Detector Blinding Attacks for Quantum Key Distribution}%

\author{Charles~Ci~Wen~Lim, Nino~Walenta, Matthieu~Legr\'{e}, Nicolas~Gisin and Hugo~Zbinden
       % <-this % stops a space
\thanks{Charles Ci Wen Lim, Nicolas Gisin and Hugo Zbinden are with Group of Applied Physics, University of Geneva, Switzerland.}% <-this % stops a space
\thanks{Nino Walenta is with Battelle, Columbus, Ohio, United States. }% <-this % stops a space
\thanks{Matthieu~Legr\'{e} is with ID Quantique SA, Geneva, Switzerland. }}

\maketitle

\begin{abstract}
In the recent decade, it has been discovered that QKD systems are extremely vulnerable to side-channel attacks. In particular, by exploiting the internal working knowledge of practical detectors, it is possible to bring them to an operating region whereby only certain target detectors are sensitive to detections. Crucially, the adversary can use this loophole to learn everything about the secret key without introducing any error to the quantum channel. In this work, as a step towards overcoming detector blinding attacks, we focus on an experimentally convenient countermeasure, where the efficiency of the detectors is randomly varied.\end{abstract}


% For peer review papers, you can put extra infomation on the cover
% \fi
\IEEEpeerreviewmaketitle

\section{Introduction}
Introduced in 1984, quantum key distribution (QKD) is a cryptographic technique that allows two spatially separated parties, called Alice and Bob, to exchange provably secure keys via an insecure quantum channel~\cite{bb84}.~Since then, much progress has been made in the theory and practice of QKD. On the theoretical side, security proofs for various QKD protocols have been obtained, and on the practical side, QKD experiments have been demonstrated under real-world conditions~\cite{Gisin2002}.~Having said that, however, there are still a few open questions to be explored before QKD can be brought onto a larger scale.~For instance, research on countermeasures against side-channel attacks is still very much a work in progress.

In the recent decade, it has been repeatedly pointed out that one can exploit the internal working knowledge of single-photon detectors to perform side-channel attacks on QKD systems~\cite{Makarov2006, Qi2006, Lydersen2010, Gerhardt2011, Jain2011}. More specifically, the adversary (called Eve) can identify or engineer detector`blind spots'' by using detailed knowledge\footnote{According to the Kerckhoffs's principle, Eve knows everything about the QKD system (e.g., the source and detectors characterization), except for the input choices (i.e., the basis choices).}~about the detectors. In other words, only selected detectors can de made sensitive, while the rest are not. This in turn provides her with a platform to concoct powerful side-channel attacks that are undetectable, at least based on known parameter estimation methods proposed in QKD. More precisely, in most QKD implementations two separate detectors are used to record bit values ``0" and ``1", respectively. Eve can either profit from the temporal mismatch of detection windows to shift the photons sent by Alice to  time intervals where one detector is sensitive and the other is insensitive, and consequently learn the bit value (time-shifting attacks~\cite{Qi2006}). Or, she can also use a so-called faked-state attack~\cite{Makarov2006}, an intercept-resend attack to learn about the secret key.

A prominent example of this latter attack is the blinding attack, which cleverly exploits the physics of single-photon detectors: Eve sends bright light pulses to bring the detectors from the Geiger mode to linear mode, where the detectors behave like classical detectors~\cite{Lydersen2010}. Then, via tailored bright pulses (henceforth called trigger pulses), Eve can control the response functions of the detectors. So for instance, she can make sure that there are only clicks if Bob uses the same measurement basis and therefore no errors are generated. Abstractly speaking, the aim is to convert the response of Bob's measurement device into one that is dependent on Eve's attack strategies.~We note that from the perspective of security proofs, blinding breaks the fundamental assumption\footnote{This assumption lies at the heart of parameter estimation:~roughly speaking, it implies that the set of quantum signals selected for parameter estimation is a representative sample of the quantum signals used to generate the secret key.} that the probability of detection is independent of the basis used to measure the incoming quantum signal~\cite{Mayers2006,Tomamichel2012aM}.~This implies that Bob's measurement data is generated from a distribution that is conditioned some variables determined by Eve; therefore, it is not astonishing that Eve can learn everything about the secret key without introducing any error. Indeed, as shown by Qin Liu~\emph{et al}~\cite{Liu2014}, the ideal control method corresponds to the case whereby the response functions of Bob's detectors is deterministic and depends only on Eve's trigger pulses and Bob's measurement choices. 

Very recently, the concept of measurement device-independent QKD (MDIQKD)~\cite{Lo2012} has been proposed as a means to overcome all possible detector side-channel attacks. This scheme is very elegant and demonstrates for the first time that quantum entanglement can be used to overcome some side-channel attacks. Roughly speaking, it employs the concept of ``time-reversal entanglement QKD" to delegate the responsibility of the detectors to an untrusted third party called Charlie, who (purportedly) performs a Bell-state-measurement on the states. In practice, however, this requires two-photon interference, which is experimentally challenging to implement~\cite{ExpMDIQKD, ExpMDIQKD2}. For this reason, it may be more practical to consider practical countermeasures that are valid against a smaller class of detector blinding attacks. For example, one may consider the countermeasure proposed in Ref.~\cite{Yuan2011} changing the detector electronics, however a security analysis has yet to be made.

In this work, we analyze the security of an experimentally convenient countermeasure against a large class of blinding attacks.~The proposal is based on an earlier idea suggested in a patent made by ID Quantique~\cite{Legre2010}, where Bob varies independently and identically the efficiency of his detectors. Note that a similar result has been obtained in Ref.~\cite{Silva2012}, where they used certain detection statistics to detect a large class of detector side-channel attacks.

\section{A countermeasure based on efficiency randomization}

 The main idea is to randomly change the parameter settings of the single-photon detectors, and then check the observed detection rates after the measurement phase.~If there is a discrepancy between the observed detection rates and the expected detection rates, Alice and Bob abort the protocol. The countermeasure can be divided into four successive steps: (1) random variation of the efficiency or activation timing (see Ref.~\cite{Legre2010} for the definition of activation timing) of the detectors (with at least 2 different values), (2) sorting of the detections by efficiency or activation timing values, (3) after sifting, comparison between the expected and measured probabilities of detection and (4) if there is a discrepancy, abort the protocol. 

Here, we try to provide the intuition behind the countermeasure from the perspectives of theory and practice.~For the latter, we briefly discuss two hacking demonstrations that have been demonstrated on a commercial QKD system called Clavis2\footnote{See http://www.idquantique.com/images/stories/PDF/clavis2-quantum-key-distribution/clavis2-specs.pdf for the specifications.}; note that QKD system is implemented with two single-photon avalanche diodes working in gated mode. This means that the detectors are in the Geiger mode when an electric gate is applied on the diodes.

The first one is called the after-gate attack~\cite{Wiechers2010}.~It consists in sending bright light (trigger) pulses slightly after the electric gate, when the detector is in the linear mode. If the photocurrent generated by the trigger pulse is sufficient to trigger the discriminator after the avalanche photodiode, a fake detection can be generated on-demand by Eve. Note that this fake detection occurs after the electrical gate, but may still be accepted by the system if the time window of the detection is larger than the gate width (which is the case in the Clavis2 QKD system). A simple countermeasure is to randomly apply a gate pulse or not (i.e. two different efficiency values unity and null). It is quite straightforward to see that Bob is able to detect the attack of Eve, since unexpected detections will be generated whenever the electrical gate is not applied. Therefore, by simply checking if the probability of detection is null when the efficiency is null, Bob can detect an after-pulse attack.

The second demonstration is the bright illumination blinding attack~\cite{Lydersen2010}, which was mentioned earlier in the introduction.~Recall that in this attack, the detectors are put into the linear mode by constant bright illumination, and this happens whenever electrical gates are applied or not. Once in the linear mode, Eve can fully control the detectors by sending in specifically tailored bright trigger pulses during the gates. Moreover, we conservatively assume that she can do it in a way that no click is generated if no gate pulse is applied (the linear gain being a bit smaller in that case). Hence, the former simple countermeasure with the two efficiency values  null and unity does not work anymore. Therefore we consider the case when the countermeasure is implemented with two efficiency values which are not null.  Suppose Bob chooses randomly between 5\% and 10\% efficiency by changing the height of the gate or the offset voltage value. The eavesdropper should be able to generate fake detections with a probability of success that depends on the efficiency value. Indeed, an attack triggering clicks with a probability smaller than 1 has been demonstrated in Ref.~\cite{Lydersen2011}. However, to our knowledge, there is no demonstration of an attack that allows Eve to generate fake detections with probabilities of success that are proportional to the quantum efficiency of the single photon detector. These examples suggest that the countermeasure  is efficient against a class of detector side-channel attacks. 

On the conceptual level, the countermeasure can be seen as a way to introduce a knowledge gap between Eve and Bob, which makes it difficult for Eve to mimic the expected conditional detection rates (i.e., conditioned on the choice of the detector efficiency).~In order to formalize the security of the countermeasure, we consider a simple blinding model inspired by bright illumination attacks, and we assume that the response functions of the blinded detectors are independent of the detector efficiency. This allows us to view the model as a deterministic function of Eve' trigger pulse and Bob's measurement choice. As we will see later, the conditional detection rates can be used to bound how often Eve does a blinding attack. In fact, incorporating this countermeasure into the asymptotic security analysis of the BB84 protocol is rather straightforward.

\section{QKD Setting}
First of all, let us define the QKD protocol and the device models which we are considering. Here, we consider a biased basis choice BB84 protocol~\cite{Lo2005A}, where one basis is chosen with higher probability than the other basis. Concerning the modeling, on Alice's side, she has a single-photon source, which she uses to prepare single-photon states in one of the four possible polarization states associated with two mutually unbiased polarization bases. Specifically, the four possible polarizations are $\ket{H},\ket{V},\ket{+},\ket{-}$, where $\{\ket{H},\ket{V}\}$ represents the linear basis and $\{\ket{+},\ket{-}\}$ represents the diagonal basis. 
 \begin{figure} \centering
  \includegraphics[width=70mm]{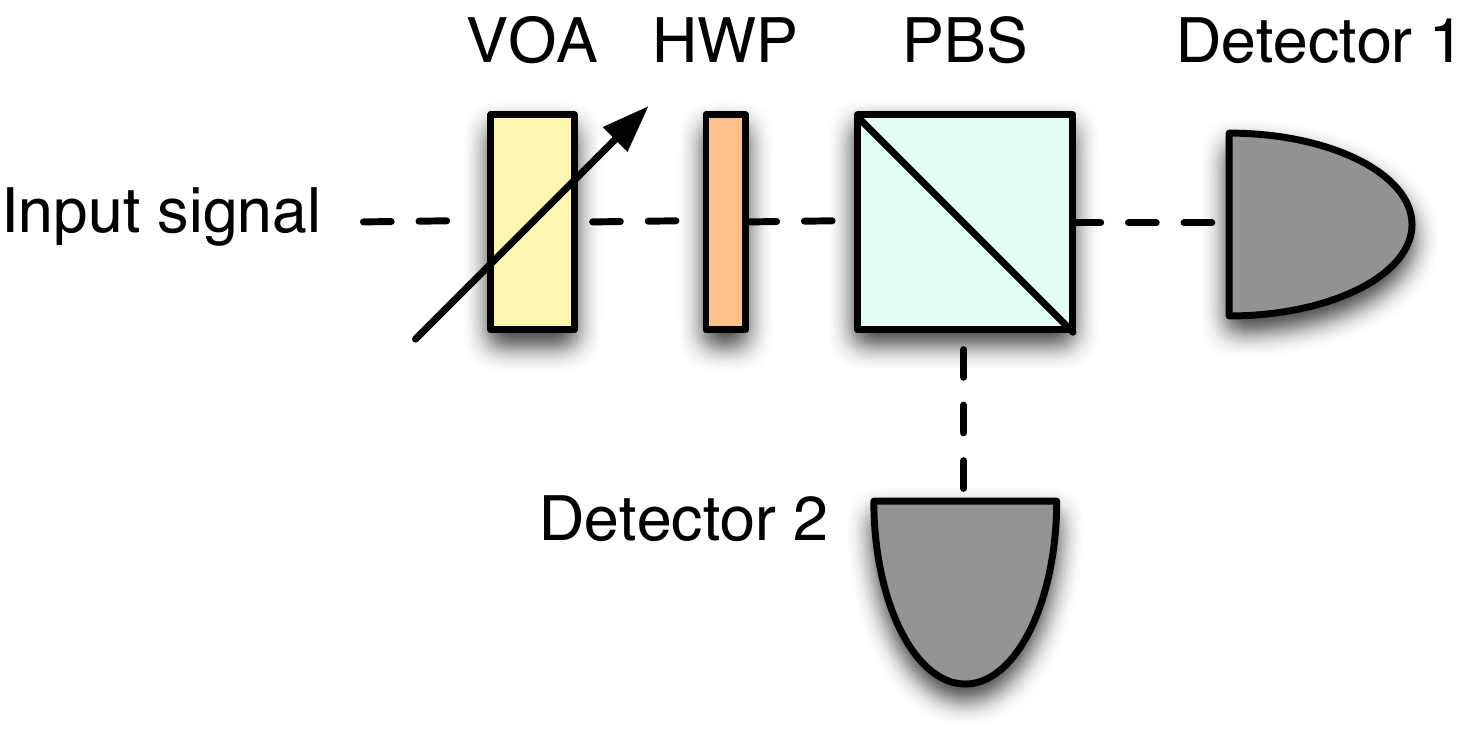} 
  \caption{{\bf Countermeasure against blinding attacks.} On the practical side, the countermeasure can be easily achieved by adding an intensity modulator (IM) in front of most existing measurement setups, or by changing the bias voltage of the detectors. The intuition behind the countermeasure can be understood by first noticing that most blinding attacks crucially rely on precise calibration of the trigger pulses. In particular, if the incident power is too low, then this would result in additional detection loss, and if the incident power is too high, there might be more double detections than expected. Importantly, we note that for security reasons, a double detection is always replaced by a random bit value~\cite{Lutkenhaus1999, Tsurumaru2008}, which in turn reduces the amount of information Eve has about Bob's bit string; likewise, this also introduces additional errors between Alice and Bob. For these reasons, Eve should send trigger pulses with a power level that maximizes her information about Bob's bit string~\cite{Liu2014}.  } 
  \end{figure}
  
  On Bob's side, he has a measurement device that allows him to measure the incoming quantum signal either in the linear basis or the diagonal basis. The measurement device is modeled using the standard active basis choice measurement setup: the basis choice is implemented using a half-wave plate and the measurement is implemented with a polarizing beam-splitter with two single-photon detectors placed at the output ports. In addition, the detectors are assumed to be ideal single-photon detectors, i.e., they have unit detection efficiency and are noiseless. Therefore, we are working in the so-called uncalibrated-device scenario.~Finally, the efficiency of the detectors is varied by using a variable attenuator. In particular, Bob can choose two levels of efficiency, $\eta_1$ and $\eta_2$, where $\eta_1 > \eta_2 \geq 0$.

The protocol basically runs as follows. \newline

{\emph{1.~Preparation.}}~First, Alice generate a long string of random bits. Then, for each bit, she prepares a single-photon (with either linear polarization or diagonal polarization) using that bit, and sends it to Bob via the quantum channel. Here, recall that the linear polarization basis is chosen with a higher probability than the diagonal basis.\newline

{\emph{2.~Measurement.}}~For each quantum signal sent by Alice, Bob randomly sets the detector efficiency (via a variable optical attenuator (VOA)) to either $\eta_1$ or $\eta_2$. In particular, he chooses $\eta_1$ most of the time. Then he measures the incoming signal in either the linear polarization basis or the diagonal basis. \newline

{\emph{3.~Sifting.}}~They publicly announce their basis choices and keep only the bits corresponding to identical basis choices. Additionally, they categorize the sifted data according to $\eta_1$ and $\eta_2$. This results in two sets of data, namely a raw key pair (using the linear polarization basis and $\eta_1$) and a parameter estimation key pair (fusing the diagonal polarization basis). \newline

{\emph{4.~Parameter Estimation.}}~They publicly announce the parameter estimation key pair to compute the error rate. In addition, conditioned on Bob's detectors efficiency choices, they can identify two sets of conditional error rates, $e_{\tn{obs},1}$ and $e_{\tn{obs},2}$, which correspond to $\eta_1$ and $\eta_2$. In additional, Bob calculates the detection rates, $R_1$ and $R_2$.\newline

{\emph{5.~Classical Post-processing.}}~Finally, they apply error correction and privacy amplification to the raw key pair to extract a secret key pair.

\section{Attack model}
Let us now introduce the attack model we are considering. We assume that (A1) Eve interacts independently and identically with each individual quantum signal, and that (A2) she either performs an attack (i.e., a blinding attack with probability $qp_c$ or a quantum attack with probability $q(1-p_c)$), or blocks the quantum channel with probability $1-q$.  Note that assumption A1 can be assumed without any loss of generality~\cite{Kraus2005}: it has been shown that it is sufficient to consider collective attacks. In the following, we specify the class of blinding attacks that we are considering.

 \begin{figure}[h!] \centering
  \includegraphics[width=70mm]{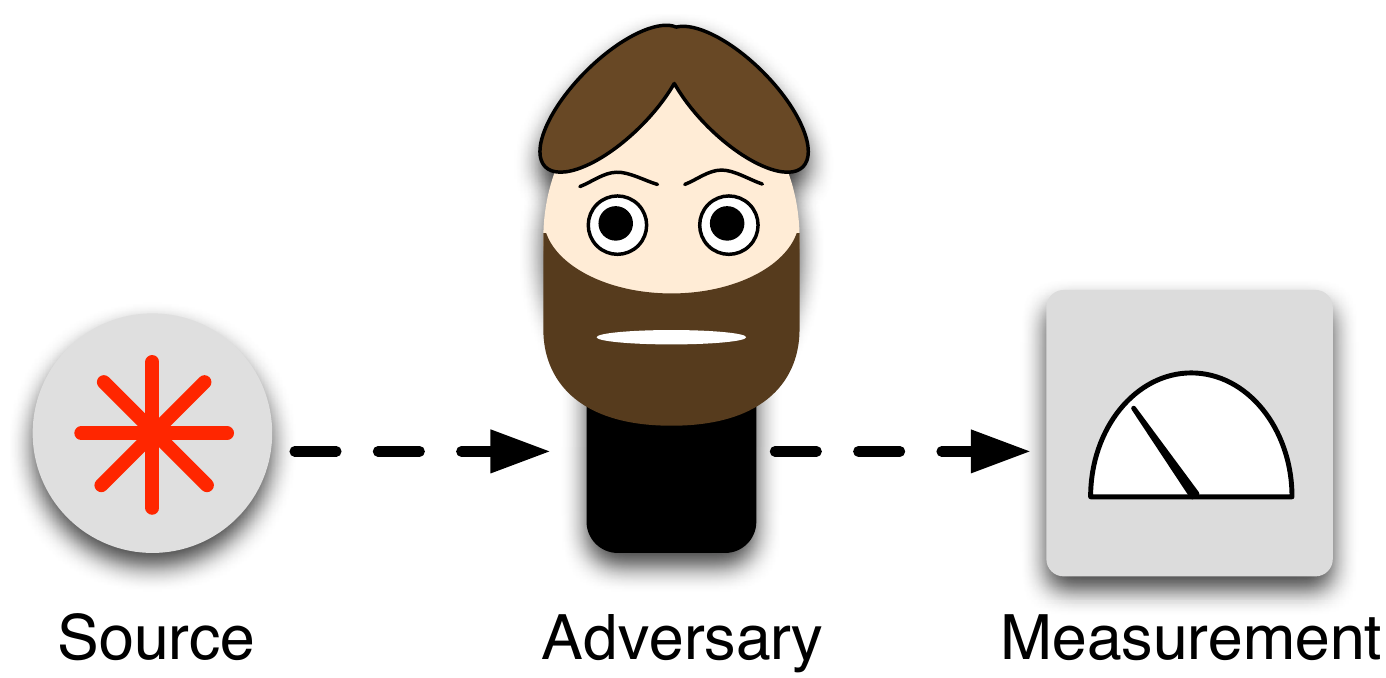} 
  \caption{{\bf Modeling attack strategies.} We consider that Eve either attacks the protocol or blocks the quantum signals. If she attacks the protocol, then she either mounts a blinding attack or a quantum attack. In the case of a blinding attack, we assume that she has complete control over the measurement device, i.e., she can externally control the response of the measurement device.   } 
  \end{figure}

\emph{Detector blinding attacks.} A blinding attack is carried out in two phases. In the first phase, Eve does the fake-state attack, i.e., she measures the quantum signal sent by Alice in one of the two possible bases and records the basis choice and outcome in $y_e$ and $b_e$, respectively. Then in the second phase, she blinds the detectors (e.g., by bright illumination) and sends a signal carrying $y_e$ and $b_e$ to the blinded measurement device. In this case, the response of the blinded measurement device is dependent on Eve's basis choice, $y_e$, and Bob's basis choice, $y$. That is, the measurement device outputs a conclusive outcome $b=b_e$ only if Bob's basis choice matches Eve's basis choice. Otherwise, the device outputs an inconclusive outcome, i.e., $b=\emptyset$. This guarantees that Eve's and Bob's bit strings are identical. %Here, it is important to mention that the measurement's response may also depend on Bob's detection efficiency choice (i.e., $\eta_1$ or $\eta_2$). However, such general blinding attacks would be 

\emph{Quantum attacks:} The quantum attacks considered here are the standard ones used in most security analyses of QKD to date, i.e., collective attacks. In addition, we assume that Eve always forward a single-photon to Bob wherever a quantum attack is mounted.

\section{Security analysis}
For simplicity, we consider the asymptotic security where Alice and Bob exchange a large number of quantum signals. This allows us to neglect all finite-size corrections necessary for the security analysis (e.g., see Ref.~\cite{Tomamichel2012aM}).

As mentioned above, the key idea of varying the efficiency of the detectors is to introduce a knowledge gap between Eve and Bob, so that the expected detection rates conditioned on the choice of detector efficiency are inaccessible to Eve. In order to formalize this intuition, we need to first relate the expected detection rates to the aforementioned attack model. The conditional detection rates are
\be
R_k=qp_c  f_c(y_e,y)+q(1-p_c)\eta_k,
\ee
for $k=1,2$. Here, $f_c(y_e,y)$ is the response function of the blinded measurement device, or equivalently, the probability of having a conclusive outcome. Note that $f_c(y_e,y)$ is upper bounded by the probability that Eve guesses correctly the basis choice of Bob. In particular, if Alice and Bob each choose the key basis (say, the linear polarization basis) with probability $p_x>1/2$, then $f_c(y_e,y) \leq 1-2p_x(1-p_x)$. Using the assumption that $f_c(y_e,y)$ is independent of Bob's detector efficiency choice, it is easy to show that
\be\label{gamma}
qp_c  f_c(y_e,y)=\max\left\{\frac{\eta_1R_2-\eta_2R_1}{\eta_1-\eta_2},0 \right \}=:\gamma.
\ee
Note that in general $f_c(y_e,y)$ may depend on the choice of $\eta_k$. However, for the type of blinding attacks considered here, it is reasonable to assume that the detector's response function is independent of $\eta_k$, since it is optimal for Eve to maximize the control of the detectors whenever they are blinded~\cite{Liu2014}.

Moreover, under the conservative assumption that blinding attacks introduce zero errors and that the errors are independent of Bob's basis choice, the probability of observing an error (conditioned on Bob choosing $\eta_k$) is
\be
g_k=q(1-p_c)\eta_k\lambda,
\ee
for $k=1,2$ and $\lambda$ is the single-photon error rate (due to quantum attacks). That is, in the asymptotic limit, Alice and Bob observe a conditional error rate of $e_{\tn{obs},k}:=g_k/R_k$. In this case, the phase error rate in the raw key pair (recall that it is formed using data from the linear basis and $\eta_1$) can be upper bounded by
\be
e_{\tn{ph}} \leq \frac{\gamma}{2R_1}+e_{\tn{obs},1}.
\ee
Indeed, if $p_c=0$, then the expected phase error rate is $e_{\tn{ph}} = \lambda$, which is simply the quantum error rate. For error correction, we only need to correct for the number of observed errors. In particular, we assume that Alice and Bob executes an one-way error correction scheme that reveals $\tn{h}_2(e_{\tn{obs},1})$ fraction of the raw key pair. Here, $\tn{h}_2(\cdot)$ is the binary entropy function. Finally, using the security bound from Ref.~\cite{SP00}, we see that the secret key fraction, i.e., the fraction of secret bits that can be extracted from the raw key pair, is 
\be \label{skr}
1-\tn{h}_2\left(\frac{\gamma}{2R_1}+e_{\tn{obs},1}
\right)-\tn{h}_2\left(e_{\tn{obs},1}\right).
\ee
From the above, we see that if $\gamma=0$, then Eq.~(\ref{skr}) becomes $1-2\tn{h}_2(e_{\tn{obs}})$, which is the secret key fraction for the BB84 protocol under the assumption of a binary symmetric channel with error rate $e_{\tn{obs}}$~\cite{SP00}.

Indeed, the intuition behind the countermeasure proposed in Ref.~\cite{Legre2010} is compactly captured in Eq.~(\ref{gamma}). Particularly, it quantifies how often Eve mounts a successful detector blinding attack in terms of the difference between the two conditional detection rates. For instance, if Eve mounts a blinding attack (that is independent of $\eta_k$), then this necessarily introduces a difference proportional to $\eta_1R_2-\eta_2R_1$. On the other hand, if Eve mounts only quantum attacks, then the expected detection rates should be $R_1=t\eta_1$ and $R_2=t\eta_2$, where $t$ is the transmissivity of the quantum channel. In this case, $\gamma=0$.  

\section{Discussion and conclusion}
As we have shown above, by adopting a simple countermeasure, it is possible to rule out a large class of powerful detector-side channel attacks. In particular, the considered countermeasure is useful against detector blinding attacks like the ones proposed in Ref.~\cite{Lydersen2010}, where Eve employs fake-states together with blinding. The central idea behind these attacks is that Eve is able to manipulate the detection efficiency of Bob's measurement device, i.e., the response of the measurement device is fully controlled by her.~Therefore, the blinded measurement device can be viewed as a black box whose response function is dependent on Eve's input signal and the measurement choice of Bob. The proposed countermeasure overcomes this by randomly varying the efficiency of the detectors, that is, under the assumption that the additional randomness is independent of Eve, the conditional detection rates can be used to reveal how often Eve makes a blinding attack. Importantly, this quantitative measure gives a bound on the phase error rate. 

Admittedly, the proposed countermeasure is not a solution to all possible detector side-channel attacks. The reason for this limitation is that our security bounds strictly requires that $f_c(y_e,y)$ is independent of $\eta_k$, which is rather restrictive and does not apply on time-shifting attacks~\cite{Qi2006}. Indeed, such attacks essentially send original quantum signals to Bob and $f_c(y_e,y)$ is dependent on $\eta_k$. However, the basic idea of randomly varying parameters of BobÕs measurement device still works, e.g., by randomly changing the time delays of the detection windows. Therefore, a combination of different convenient countermeasures have to be implemented in order to rule out a large class of side-channel attacks.

\section*{Acknowledgment}
We would like to thank Nitin Jain for his suggestions and comments. Additionally, we would like to thank EU Marie Curie IAPP project QCERT and the Swiss NCCR QSIT for financial support. ID Quantique would like to thank the European Union project SIQS and EMRP (project IND06-MIQC). The EMRP is jointly funded by the EMRP participating countries within EURAMET and the European Union.

\ifCLASSOPTIONcaptionsoff
  \newpage
\fi


\begin{thebibliography}{99}
\bibliographystyle{unsrt}


\bibitem{bb84}
C. H. Bennett and G. Brassard, 
%\emph{Quantum Cryptography: Public key distribution and Coin tossing},
\emph{Proceedings of IEEE International Conference on Computers, Systems, and Signal Processing}, 
Bangalore, India, IEEE Press (New York), 1984, pp. 175-179.

\bibitem{Gisin2002}
N. Gisin, G. Ribordy, W. Tittel and H. Zbinden, %\emph{Quantum cryptography},
Rev. Mod. Phys. {\textbf{74}}, 145-195 (2002);
V. Scarani \emph{et al.}, %H. Bechmann-Pasquinucci, N. J. Cerf, M. Du\v{s}ek, N. L\"{u}tkenhaus and M. Peev,
%\emph{The security of practical quantum key distribution},
Rev. Mod. Phys. {\textbf{81}}, 1301-1350 (2009).

\bibitem{Makarov2006}
V. Makarov, A. Anisimov and Johannes Skaar, 
%\newblock {\em Effects of detector efficiency mismatch on security of quantum cryptosystems},
Phys. Rev. A. {\textbf{574}}, 022313 (2006).

\bibitem{Qi2006}
B. Qi, C.-H. F. Fung, H.-K. Lo and X. Ma,
Quantum Inf. Comput. \textbf{7}, 073 - 082 (2007).

\bibitem{Lydersen2010}
L. Lydersen \emph{et al.},
Nature Photon. \textbf{4}, 686-689, (2010).

\bibitem{Gerhardt2011}
I. Gerhardt, Q. Liu, A. Lamas-Linares, J. Skaar, C. Kurtsiefer and V. Makarov,
Nat. Commn. \textbf{2}, 349 (2011).

\bibitem{Jain2011}
N. Jain, Christoffer Wittmann, Lars Lydersen, \emph{et al},
Phys. Rev. Lett. {\bf{107}}, 110501 (2011).

\bibitem{Tomamichel2012aM}
M. Tomamichel, C. C. W. Lim, N. Gisin and R. Renner,
%\emph{Tight finite-key analysis for quantum cryptography},
Nature Commun. {\textbf{3}}, 634 (2012).

\bibitem{Mayers2006}
H. Inamori, N. L\"{u}tkenhaus and D. Mayers,
The European Physical Journal D. \textbf{41}, 3, (2007).

\bibitem{Liu2014}
Q. Liu, A. Lamas-Linares, 
C. Kurtsiefer, J. Skaar, V. Makarov and I. Gerhardt
Rev. Sci. Instrum. \textbf{85}, 013108, (2014).

 \bibitem{Lo2012}
H.-K. Lo, M. Curty and B. Qi,
Phys. Rev. Lett. {\bf{108}}, 130503 (2012).

\bibitem{ExpMDIQKD}
A. Rubenok, J. A. Slater, P. Chan, I. Lucio-Martinez and W. Tittel, Phys. Rev. Lett. {\textbf{111}}, 130501 (2013)

\bibitem{ExpMDIQKD2}
T. Ferreira da Silva, D. Vitoreti, G. B. Xavier, G. P. Tempor\~{a}o and J. von der Weid, Phys. Rev. A. {\textbf{88}}, 052303 (2013).
%
 \bibitem{Yuan2011}
 Z. L. Yuan, J. F. Dynes and A. J. Shields,
 Appl. Phys. Lett. {\bf{92}}, 231104 (2011).
 

 \bibitem{Legre2010}
M. Legr\'{e} and G. Ribordy, intl. patent. appl. WO 2012/046135 A2 (filed in 2010).

\bibitem{Silva2012}
T. Ferreira da Silva, G. B. Xavier, G. P. Tempor\~{a}o and J. von der Weid, 
Opt. Express. {\bf{20}}, 17, 18911-18924 (2012)

\bibitem{Wiechers2010}
C. Wiechers, L. Lydersen, C. Wittmann, D. Elser, J. Skaar, C. Marquardt, V. Makarov and G. Leuchs,
New J. Phys. \textbf{13}, 013043 (2011).

\bibitem{Lydersen2011}
L. Lydersen, N. Jain, C. Wittmann, \emph{et al},
Phys. Rev. A \textbf{84}, 032320 (2011).


\bibitem{Lutkenhaus1999}
N. L\"{u}tkenhaus,
Phys. Rev. A. \textbf{59}, 3301 (1999).

\bibitem{Tsurumaru2008}
T. Tsurumaru and K. Tamaki,
Phys. Rev. A. \textbf{78}, 032302 (2008).


\bibitem{Lo2005A}
H.-K. Lo, H. F. Chau and M. Ardehali, 
J. Cryptology {\textbf{18}}, No. 2, 133 (2005).


\bibitem{Kraus2005}
B. Kraus, N. Gisin and R. Renner
%\newblock {\em Quantum cryptography based on Bell's theorem}.
Phys. Rev. Lett. {\textbf{95}}, 080501 (2005).

\bibitem{SP00}
P. W. Shor and J. Preskill,
%\newblock {\em Simple Proof of Security of the BB84 Quantum Key Distribution
  %Protocol},
Phys. Rev. Lett. {\textbf{85}}, 441 (2000).

\end{thebibliography}
\end{document}